# PoDT: A Secure Multi-chains Consensus Scheme Against Diverse Miners Behaviors Attacks in Blockchain Networks

Wenbo Zhang, Tao Wang, Jingyu Feng*

*Abstract*—As cross-chain technologies make the interactions among different blockchains (hereinafter "chains") possible, multi-chains consensus is becoming more and more important in blockchain networks. However, more attention has been paid to the single-chain consensus schemes. The multi-chains consensus with trusted miners participation has been not considered, thus offering opportunities for malicious users to launch Diverse Miners Behaviors (DMB) attacks on different chains. DMB attackers can be friendly in the consensus process of some chains called mask-chains to enhance trust value, while on other chains called kill-chains they engage in destructive behaviors of network. In this paper, we propose a multi-chains consensus scheme named as Proof-of-DiscTrust (PoDT) to defend against DMB attacks. Distinctive trust idea (DiscTrust) is introduced to evaluate the trust value of each user concerning different chains. A dynamic behaviors prediction scheme is designed to strengthen DiscTrust to prevent intensive DMB attackers who maintain high trust by alternately creating true or false blocks on kill-chains. On this basis, a trusted miners selection algorithm for multi-chains can be achieved at a round of block creation. Experimental results show that PoDT is secure against DMB attacks and more effective than traditional consensus schemes in multi-chains environments.

## I. INTRODUCTION

Blockchain, originally devised for Bitcoin [1], has evolved into a well-studied status both in industry and academia. It has an impact on all sectors of the industry and becomes a revolution [2]. The reason why blockchain can get so much attention is that it has numerous benefits [3], such as decentralization, persistence, pseudonymity, and auditability. Blockchain has been applied to many fields, including financial services [4], medicine [5], the internet of things [6], intelligent transportation system [7], e-government [8], smart advertising network [9], and so on.

Currently, cross-chain [10] has gradually become the focus, which aims to build reliable interaction channels between different blockchains (hereinafter "chains"). Some cross-chain technologies have appeared in [12-15]. They connect the dispersed blockchain ecological islands and have become a bridge link for the overall expansion of blockchain. While cross-chain facilitates the interactions among different chains, it also enables multi-chains consensus increasingly important in blockchain networks.

One of the cores blockchain technologies is consensus scheme. In blockchain networks, blocks can be validated, shared, synchronized and created across users via a peer-to-peer decentralized consensus scheme [15]. The users responsible for creating blocks in consensus scheme are called miners. Many efforts have been made to some consensus schemes [16–19], in which more attention has been paid to the consensus scheme with fair multi-miners participation. For the blockchain platform, one of the most important properties is security [20]. At present, most trust management mechanisms mainly focus on the enhancement of the whole network's security or robustness [21]. To detect malicious users, trust management can be introduced to estimate whether a user is honest or not by his historical behaviors, and thus selecting trust miners in the light of their trust value to participate in the consensus scheme.

However, most of the consensus schemes are based on single-chain mode and universal trust evaluation. This may offer opportunities for malicious users in the multi-chains consensus scheme. When there are multiple chains in a blockchain network, some miners may exist on all chains so that they can handle different businesses in different blockchains at the same time, which can help them grasp all information or dispatch different tasks. Once these users are attacked or even hijacked, it is possible for malicious users to do evil on a chain if the high trust value generated by their honest behaviors on another chain is universal to the whole network. That is, malicious users may launch Diverse Miners Behaviors (DMB) attacks on different chains.

In this paper, we propose a multi-chains consensus scheme called Proof-of-DiscTrust (PoDT) along with a proof-of-concept blockchain design to defend against DMB attacks. The main contributions of this paper are as follows.

- Conduct an in-depth investigation on DMB attacks. According to the strategy of DMB attacks, the chains in the network can be divided into kill-chains and mask-chains. The basic idea of DMB is as follows. DMB attackers may behave diversely on different chains. They can be friendly in the consensus process of mask-chains to enhance their trust value and remain honest miners, while in kill-chains they engage in behaviors that undermine the consensus process. Specifically, normal DMB attackers exploit the high trust value achieved from mask-chains to engage in sabotage on kill-chains. Further, intensive DMB attackers maintain high trust value by alternately creating true or false blocks on kill-chains, which would make them harder to detect in the consensus scheme.
- Introduce the distinctive trust idea (DiscTrust) to evaluate the trust value of each user concerning different chains. The trustworthiness of a user is split into the local and global trust value. The evaluation of local trust value for a user is bound to each chain. The higher local trust value of a user just shows he is trustworthy on a chain rather than all chains. Only when the local trust value concerning all chains and the global trust value are all higher will the user be recognized as honest. In this case, normal DMB attackers can be detected due to their lower local trust value on kill-chains.
- Propose a Dynamic Behaviors Prediction (DBP) scheme on the basis of DiscTrust. An additional side chain is employed to store the historical experiences of users on all chains. Considering that SVM is highly accurate and fast in dichotomy prediction, it can be well applied to DBP. By analyzing the historical experiences of a user, the prediction model can be built by combining the Lagrangian multiplier into an objective function to predict his dynamic behaviors on a kill-chain. If the prediction result is +1, intensive DMB attackers can be detected.
- Design three algorithms to select trusted miners for multi-chains. The global trust value is used to design the algorithm of trusted network miners selection. Only after a user has been selected as a network miner will he be eligible to be selected as a chain miner who has the authority to create blocks in the consensus scheme. Chain miners who can be trusted on one chain may not be trusted on another. Thus, both the local trust value and DBP are used to design the algorithm of trusted chain miners selection on a chain. To confirm the block validation, the last algorithm is designed to elect the chain miner leader periodically.

The organization of this paper is as follows. Section II introduces the related works on cross-chain, consensus schemes and trust management. In section III, we analyze DMB attacks. The consensus PoDT scheme is proposed in Section IV to suppress DMB attacks in multi-chains environments. The performance analysis is conducted in Section V. We discuss the industrial applications of our PoDT scheme in Section VI. Finally, we conclude this paper with some future works in Section VII.

## II. RELATED WORKS

### A. Cross-chain

The emergence of blockchain has promoted the development of many industries, but a limitation may arise because of lacking interoperability between multiple chains. Fortunately, cross-chain can break the mutual isolation between chains. By realizing the internet of chain, cross-chain can expand the application scope of blockchain. Thus it can be seen that the multi-chains symbiosis has become the potential trend of blockchain networks.

Recently, some cross-chain technologies have been proposed. In [11], the blockchain router is introduced to empower chains to connect and communicate cross chains. In [12], the atomic cross-chain swap is modeled as a directed graph, whose vertexes are parties and whose arcs are proposed asset transfers. In [13], the DeXTT cross-chain transfer protocol is proposed, which can be used to record a token transfer in any number of blockchains simultaneously in a decentralized manner. In [14], it proposed a joint operation mechanism of cross-chain trading, combined distributed photovoltaic power generation market and the carbon market by the blockchain technology. In [15], cross-chain is applied to inventory complementation among enterprises to discover the shortage of inventory in time and prevent the unnecessary loss of enterprises.

Despite cross-chain broadens the application scenarios of blockchain, it also brings some technical challenges. For example, malicious users are more likely to take advantage of the multi-chains context to make adverse actions to undermine the consensus scheme.

### B. Consensus Schemes

As the core of a blockchain network, consensus schemes can guarantee that the next block of the network is the only version of the truth, and protect the network from adversarial influences on the users and the network [22-23].

The first consensus scheme is PoW [1], which was initially used to defend against 51% of attacks through the solution of puzzles. Nevertheless, PoW requires a great deal of computational power to resolve the puzzle, which results in high energy and computing resources consumption. It may serve a good purpose on Bitcoin network, but cannot meet the needs of other chains that require creating block rapidly. An improved version of PoW named GSCS has been proposed in [24], GSCS has the potential to ensure a more secure and robust environment for decentralized blockchain systems.

Proof-of-stake (PoS) [16] concept states that a person can mine or validate block transactions according to how many coins he or she holds instead of resolving a puzzle. Although PoS can save computational power effectively, it will make the rich get richer. In [25], an improved consensus algorithm named Delegated Proof of Stake with Downgrade (DDPoS) is proposed, which can detect and downgrade the malicious nodes timely to ensure the security and good operation of the system. A behavior-based incentive mechanism named Proof-of-Behavior (PoB) is introduced in [26], which can stimulate honest behavior and neutralize malicious attacks.

In recent years, more attention has been paid to the consensus schemes with fair multi-miners participation. Each user can fairly compete to become a miner in a blockchain network, rather than relying on computational power or coins. For instance, PBFT [17] has been widely noted for allowing all users of the blockchain to participate in consensus, and for being able to withstand up to one-third of malicious attacks. But it can only prevent malicious users from voting maliciously and false blocks from being created on the single chain. Moreover, these malicious users cannot be accurately detected by PBFT. Tendermint [18] is also a classical BFT (Byzantine Fault Tolerance) consensus scheme that can work even if up to one-third of users in the network fail in arbitrary ways. Based on Tendermint, a consensus scheme that exploits randomness and game theory is proposed in [19], but the trust evaluation of the random miners is not considered on the block creation.

Although there has been a lot of research works on consensus mechanism, most of them are based on single-chain mode. The multi-chain consensus scheme with the ability to identify malicious users has not yet been developed.

### C. Trust Management

Trust management has been increasing influence on many application scenarios, including e-commerce [27], online social communities [28], wireless sensor networks [29], and so on.

To prevent malicious users from joining the miners team, trust management should be introduced. In blockchain networks, representative trust management systems are as follows.

In [30]，a trustworthiness calculation method based on trust blockchain users is proposed. In [31], TrustChain is proposed as a three-layered trust management framework which uses a consortium blockchain to track interactions among supply chain participants and to dynamically assign trust and reputation scores based on these interactions. In [32], a blockchain based trust mechanism is presented, in which the edge reputation system chooses the miner of the blockchain for the joint Proof-of-Stake consensus protocol to append a block recording the new service reputations. In [33], a consensus scheme with fair multi-miners participation called PoN is proposed, in which the trust evaluation is designed to select trusted miners in the single-chain consensus scheme.

Currently, one of the most popular trust management designs is based on the beta function. It first counts the number of honest and dishonest behaviors a user has conducted, and then calculates the trust value with beta function denoted by $Beta(\alpha, \beta)$ [34].

$$Beta(\alpha, \beta) = \frac{\Gamma(\alpha, \beta)}{\Gamma(\alpha)\Gamma(\beta)} \delta^{\alpha-1}(1-\delta)^{\beta-1}, \quad (1)$$

where $\delta$ is the probability of user's behaviors, $0<\delta<1$, $\alpha>0$, $\beta>0$.

A basic trust management system called Baseline can be adopted to evaluate trust value in blockchain networks. For a user $U_i$, the system calculates the number of true blocks created by $U_i$, with the role of a miner, denoted by $tru_i$, and the number of false blocks created by $U_i$, denoted by $fal_i$. When the blocks created by $U_i$ are trusted, $\alpha=tru_i+1$, otherwise $\beta=fal_i+1$. Thus, the trust value of $U_i$ is calculated with beta function as:

$$t_i = Beta(tru_i+1, fal_i+1). \quad (2)$$

In addition, consider the condition $\Gamma(x)=(x-1)!$ when $x$ is an integer [35]. It can be found that the expectation value of the beta function is denoted as $E[Beta(\alpha, \beta)]=\alpha/(\alpha+\beta)$. In order for the initial trust value of $U_i$ to be the threshold $\theta$, the evaluation of $t_i$ can be further described as:

$$t_i = \frac{tru_i + \theta}{tru_i + fal_i + \theta}. \quad (3)$$

In the single-chain consensus scheme, current trust management can make ordinary attacks more difficult to succeed. But in the multi-chains consensus scheme, this successful foundation is built on the fact that ordinary attackers always create false blocks on all chains. To avoid detection, malicious users may strategically create false blocks. How to select trusted miners is becoming a key issue in the multi-chains consensus scheme.

## III. DMB ATTACKS OVERVIEW

When the trust management system is employed in blockchain networks, the false block threats can be easily suppressed if ordinary attackers always create false blocks against consensus scheme. This is because they will obtain a lower trust value when they always create false blocks. To avoid the detection of trust management, attackers will change their strategies.

Due to the advent of multi-chains, a new attack chance may be offered to attackers. We find that DMB attacks are applicable under three key factors: 1) each user in a blockchain network has the opportunity to act as a miner, 2) the traditional single-chain thinking mode makes the trust value universal in the whole blockchain network, and 3) no measures have been adopted to evaluate the trust value of each user concerning different chains.

In short, ordinary attackers would break the consensus scheme of all chains, while DMB attackers may exhibit diverse consensus behaviors on different chains.

Generally speaking, DMB attackers divide the chains in the network into kill-chains and mask-chains to implement their attack strategy.

• **Kill-chains**: The kill-chains are the chains whose consensus scheme is the destruction target of DMB attackers.

• **Mask-chains**: The mask-chains are the chains where DMB attackers can be friendly in the consensus process to enhance their trust value. By disguising as honest miners, they can undermine the consensus process of kill-chains.

DMB attacks can be launched from two aspects of threats: normal and intensive. For the normal DMB attackers, they always behave viciously on kill-chains. For the intensive DMB attackers, they would maintain high trust by alternately creating true or false blocks on kill-chains.

DMB attackers are extremely sensitive to their trust value. They begin to launch DMB attacks under the constraint

$$\begin{cases} gt_i \leq \theta + \xi_1 \xrightarrow{\text{launch}} \text{normal DMB attack,} \\ lt_{ij} \leq \theta + \xi_1 \xrightarrow{\text{launch}} \text{intensive DMB attack.} \end{cases} \quad (4)$$

The strategy of DMB attacks is shown in Fig. 1. Assuming $U_i$ is one of the DMB attackers, $gt_i$ is the global trust value of $U_i$ and $lt_{ij}$ is the local trust value of $U_i$ corresponding to the $j$-th chain (Chain$_j$). As each $gt_i \in [0,1]$ or $lt_{ij} \in [0,1]$, the threshold of trust value ($\theta$) is usually set to a moderate value, such as 0.5, which can be calculated in equation (1) when $tru=fal$. When $gt_i \leq \theta + \xi_1$, $U_i$ will launch normal DMB attacks to enhance his global trust value on mask-chains. Here, $\xi_1$ is the trust value warning line. It is too late to increase trust value when $gt_i \leq \theta$. This attack pattern will continue until $gt_i \geq \theta + \xi_2$ ($\xi_2$ is the high trust line). During $\theta + \xi_1 \leq gt_i \leq \theta + \xi_2$, $U_i$ will launch normal DMB attacks to undermine the consensus process of kill-chains since $U_i$ can disguise as an honest miner.

If $lt_{ij}$ is introduced, $U_i$ cannot engage in sabotage on the kill-chains with intensive DMB attacks. $U_i$ may get crafty and adopt normal DMB attacks to get the destructive chance on kill-chains. Assuming Chain$_j$ is a kill-chain, $U_i$ will launch intensive DMB attacks to enhance his local trust value on Chain$_j$ when $lt_{ij} \leq \theta + \xi_1$. This attack pattern will continue until $lt_{ij} \geq \theta + \xi_2$. During $\theta + \xi_1 \leq lt_{ij} \leq \theta + \xi_2$, $U_i$ will launch intensive DMB attacks to undermine the consensus process of Chain$_j$.

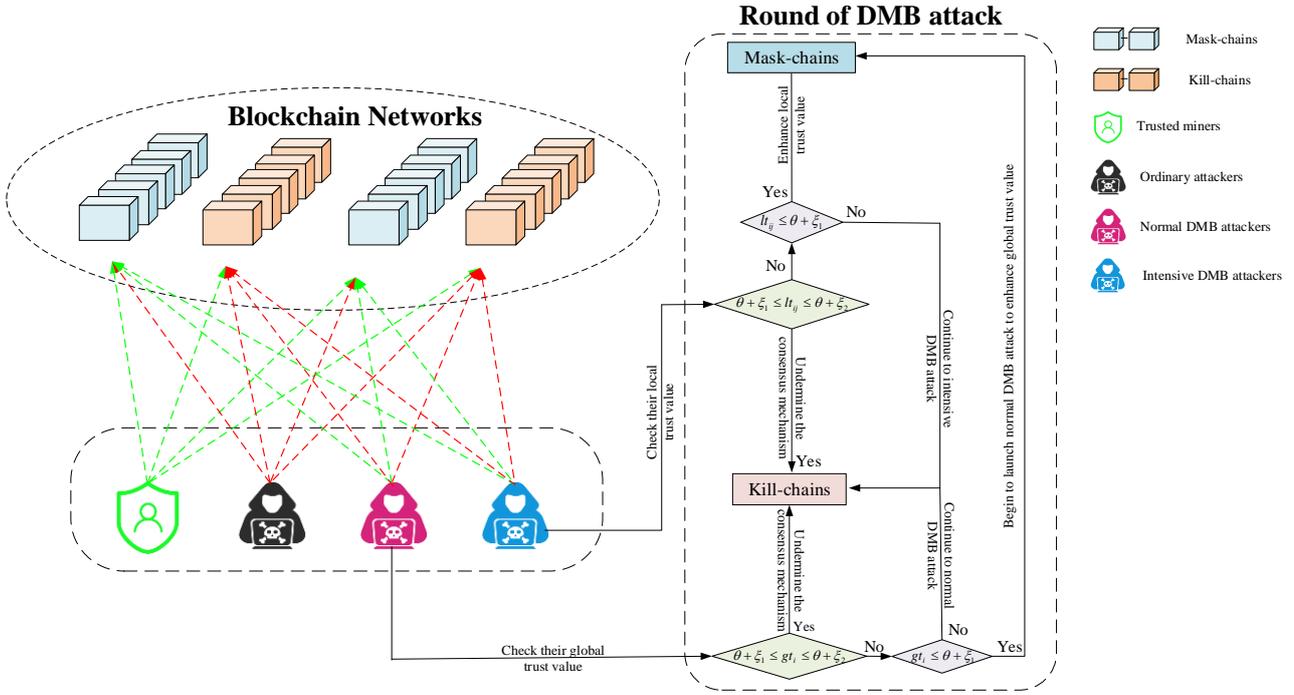

Fig. 1. Strategy of DMB attacks.

## IV. OUR PROPOSED PODT SCHEME

To select trusted miners in multi-chains, we propose a multi-chains consensus scheme called PoDT under the guidance of Proof-of-DiscTrust.

### A. Design Idea

Some negotiation rules are necessary to achieve a fast and efficient consensus scheme in a distributed manner [33]. To construct a multi-chains consensus scheme better, the following negotiation rules should be involved.

- **Rule 1:** The global trust value is used to select and update network miners ($\Theta_{net}$), while the local trust value is used to select and update chain miners ($\Theta_{chain}$).
- **Rule 2:** The size of members $|\Theta_{net}| > n/2$ where $n$ is the number of users in the network.
- **Rule 3:** The size of members $|\Theta_{chain}| > m/2$ where $m$ is the maximum number of active users in a blockchain.
- **Rule 4:** Only after a user has been selected as a network miner will he be eligible to be selected as a chain miner.
- **Rule 5:** Only chain miners have the authority to create blocks in the consensus scheme.
- **Rule 6:** Chain miners who can be trusted on one blockchain may not be trusted on another.
- **Rule 7:** A round of block creation is usually made up of block generation, block validation and block acceptance.
- **Rule 8:** In a round of block creation, several chain miners are first randomly selected to be responsible for block generation, and then several chain miners are randomly selected to perform block validation.
- **Rule 9:** On each blockchain, the chain miner leader shall be elected periodically to accept and broadcast the confirmation of block validation.

With the negotiation rules, the architectural view of PoDT is shown in Fig. 2. To suppress DMB attacks, we design PoDT from the design idea of two-level defense.

In the first-level defense, DiscTrust is introduced to defend against normal DMB attacks. DiscTrust divides a user's trust value into global trust value and local trust value. Although the global trust value of normal DMB attackers is higher than the threshold, their local trust value in kill-chains will be lower than the threshold. Therefore, DiscTrust can effectively detect normal DMB attackers. If the global trust value is less than the threshold, DiscTrust can also be used to detect ordinary attackers.

Nevertheless, the local and global trust value of intensive DMB attackers are both higher than the threshold, so they may not be detected. Supported by DiscTrust, the second-level defense can collect the dynamic changes of local and global trust value of users and propose the Dynamic Behaviors Prediction (DBP) scheme to detect intensive DMB attackers, in order to prevent kill-chains from falling their victims.

On this basis, trusted miners selection in multi-chains can be achieved at a round of block creation

### B. Distinctive Trust Evaluation

We have known that normal DMB attackers fake blocks on kill-chains and create true blocks to boost their trust value on mask-chains. In the DiscTrust scheme, our design idea is that the trust value of each user should be distinctively evaluated by his past behaviors concerning various chains respectively. As shown in Fig. 3, the DiscTrust scheme is built with three functional modules, including local distinctive trust evaluation, global distinctive trust evaluation and normal DMB attackers.

#### 1) Local Distinctive Trust Evaluation

In the DiscTrust scheme, the result of trust evaluation for various chains is not a signal value, but a set of local trust values. Take the $i$-th user ($U_i$) as an example, the local trust value of $U_i$ corresponding to the $j$-th blockchain ($Chain_j$) can be evaluated as:

$$lt_{ij} = \frac{tru_{ij} + \theta}{tru_{ij} + fal_{ij} + 1}, \quad (5)$$

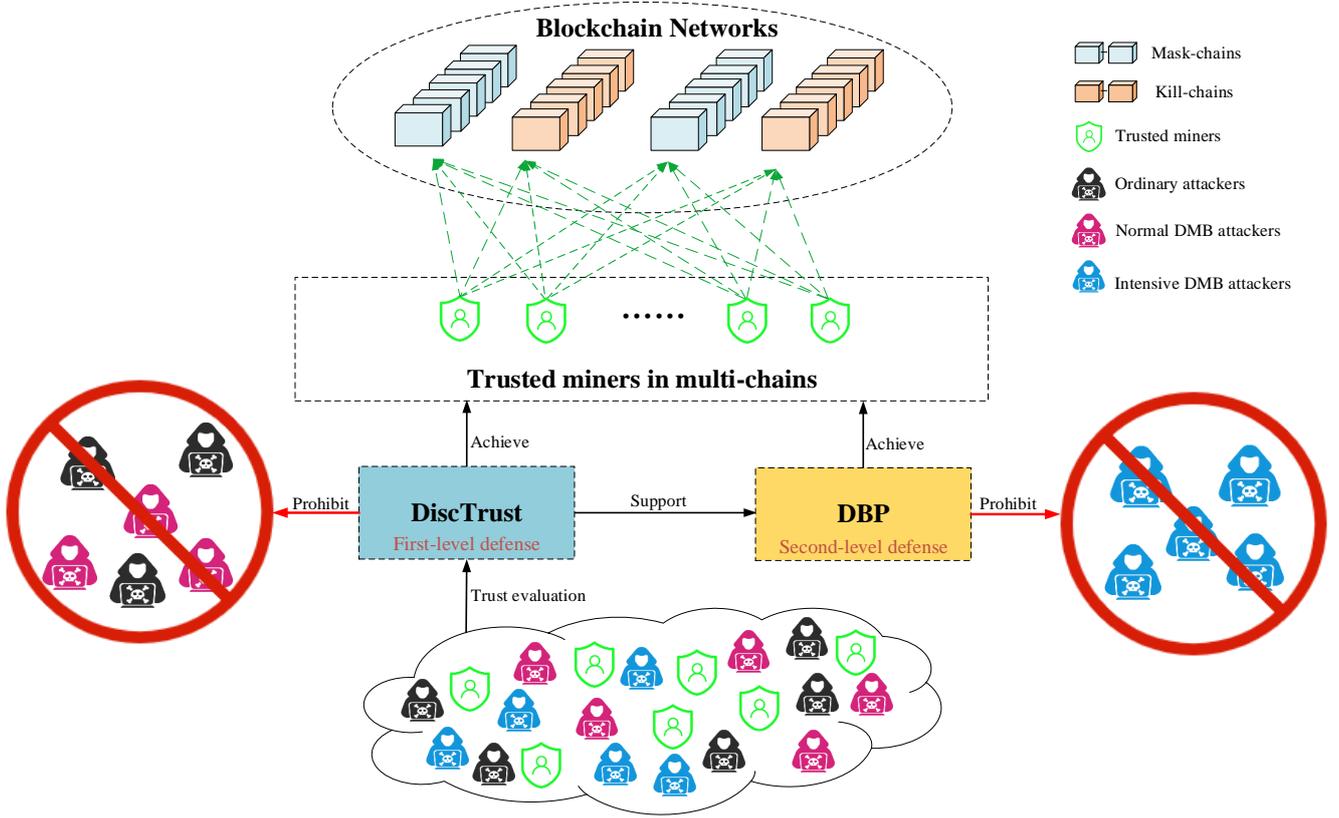

Fig. 2. Architectural view of PoDT

where the dual parameters ($tru_{ij}$, $fal_{ij}$) denote the numbers of honest and false blocks created by $U_i$ for $Chain_j$, respectively.

When $tru_{ij}=fal_{ij}=0$, $U_i$ is a newcomer and his local trust value is set to the threshold. Thus, the newcomer would have a chance to participate in the consensus scheme of $Chain_j$, and then his local trust value will be changed in accordance with his future behaviors.

Similarly, we can evaluate the local trust value of $U_i$ for all chains in the network, and thus generating a trust vector for $U_i$, which can be expressed as:

$$LT_i = [lt_{i1},\cdots,lt_{ij},\cdots,lt_{ih}]. \quad (6)$$

In a blockchain network, the trust vectors of all users can compose a matrix $LT_{n\times h}$, where $n$ is the number of users and $h$ is the number of chains in the network.

$$LT_{n\times h} = \begin{bmatrix} lt_{11} & \cdots & lt_{1j} & \cdots & lt_{1h} \\ \vdots & \cdots & \vdots & \cdots & \vdots \\ lt_{i1} & \cdots & lt_{ij} & \cdots & lt_{ih} \\ \vdots & \cdots & \vdots & \cdots & \vdots \\ lt_{n1} & \cdots & lt_{nj} & \cdots & lt_{nh} \end{bmatrix}. \quad (7)$$

*2) Global Distinctive Trust Evaluation*

To analyze the holistic behaviors of a user for all chains, the global trust value of the user corresponding to all chains should be taken into account. Take $U_i$ as an example again, the global trust value of $U_i$ corresponding to all chains can be evaluated as:

$$gt_i = \frac{tru_i + \theta}{tru_i + fal_i + 1}, \quad (8)$$

where $tru_i = \sum_{j=1}^{h} tru_{ij}$ and $fal_i = \sum_{j=1}^{h} fal_{ij}$.

For all users, $\Xi$ denotes the set of their global trust values, that are extremely valuable for trusted network miners selection.

*3) Normal DMB Attackers Detection*

To detect normal DMB attackers effectively, the trust state division of a user should be considered through the global trust value and the amount of low local trust values ($\lambda$).

$U_i$, $\lambda_i$ can be counted by our proposed Algorithm 1.

---

**Algorithm 1** Count $\lambda_i$

**Input:** $LT_i$

**Output:** $\lambda_i$

1: Initialize $\lambda_i = 0$
2: **for** each $lt_{ij} \in LT_i$ ($1 \leq i \leq h$) **do**
3:    **if** ($lt_{ij} < \theta$) **then**
4:       $\lambda_i = \lambda_i + 1$
5:    **end if**
6: **end for**

---

With ($gt_i$, $\lambda_i$), the trust state of $U_i$ can be divided into four categories.

◇ **Trustworthy state ($gt_i \geq \theta$ && $\lambda_i = 0$)**

This state shows that $U_i$ always creates true blocks for all chains. His blocks can be accepted in the current chain.

◇ **Low-risk state ($gt_i \geq \theta$ && $\lambda_i \geq 1$)**

This state shows that $U_i$ creates false blocks for a small number of chains. $U_i$ must be rejected to participate in the consensus scheme of $Chain_j$ for $lt_{ij} < \theta$.

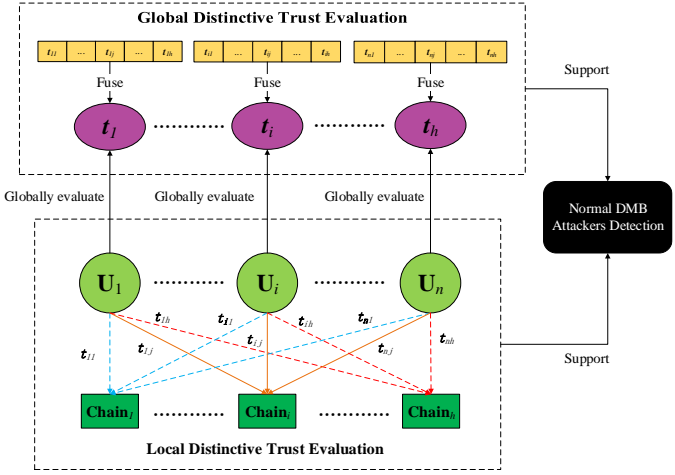

Fig. 3. Functional modules in the DiscTrust scheme.

◇ **Medium-risk state ($gt_i<\theta$ && $\lambda_i\geq1$)**

This state shows that $U_i$ creates false blocks for most of chains. $U_i$ must be rejected to participate in the consensus scheme of $Chain_j$ for $lt_{ij}<\theta$. Meanwhile, his blocks should be rejected since his trust state might be converted to the high-risk state.

◇ **High-risk state ($gt_i<\theta$ && $\lambda_i=h$)**

This state shows that $U_i$ always creates false blocks for all chains. His blocks must be rejected in the current chain.

Based on the four categories of trust state, Algorithm 2 is designed to separate the set of users ($\Phi$) into four clusters to detect normal DMB attackers in the current chain.

---

**Algorithm 2** Our proposed Normal DMB attackers detection

**Input:** $\Phi$

**Output:** $\Phi_1$, $\Phi_2$, $\Phi_3$, $\Phi_4$

1: Initialize $\Phi_1=\Phi_2=\Phi_3=\Phi_4=\varnothing$
2: **for** each $U_i\in\Phi$ **do**
3:    Observe $gt_i$ and $\lambda_i$
4:    **if** ($gt_i\geq\theta$ && $\lambda_i=0$) **then**
5:       $\Phi_1=\{U_i\}\cup\Phi_1$
6:    **elseif** ($gt_i\geq\theta$ && $\lambda_i\geq1$) **then**
7:       $\Phi_2=\{U_i\}\cup\Phi_2$
8:    **elseif** ($gt_i<\theta$ && $\lambda_i\geq1$) **then**
9:       $\Phi_3=\{U_i\}\cup\Phi_3$
10:   **elseif** ($gt_i<\theta$ && $\lambda_i=h$) **then**
11:      $\Phi_4=\{U_i\}\cup\Phi_4$
12:   **end if**
13: **end for**

---

Obviously, both $\Phi_2$ and $\Phi_3$ are the set of normal DMB attackers. We can also find that $\Phi_1$ is the set of honest miners and $\Phi_4$ is the set of ordinary attackers who always create false blocks on all chains.

*C. Dynamic Behaviors Prediction*

Intensive DMB attackers maintain high trust by alternately creating true or false blocks on kill-chains, which may allow them to hide in $\Phi_1$. By analyzing the historical experiences of $U_i$, the Dynamic Behaviors Prediction (DBP) scheme can predict his dynamic behaviors on a kill-chain such as $Chain_j$, and thus detecting whether $U_i$ is an intensive DMB attacker.

As shown in Fig. 4, on account of the function that blockchain can securely maintains transactions and records[36], an additional side chain is employed to store the historical experiences of users for all chains. The historical experiences stored in the side chain is shown in Table I.

TABLE I
HISTORICAL EXPERIENCE

| Parameters | Description |
|---|---|
| $lt_{ij}$ | Local trust value of $U_i$ corresponding to $Chain_j$ |
| $gt_i$ | Global trust value of $U_i$ |
| $t_i$ | Number of true blocks created by $U_i$ |
| $f_i$ | Number of false blocks created by $U_i$ |
| $L_j$ | Length of $Chain_j$ |
| $N_j$ | Number of active users on $Chain_j$ |
| $F^k_{ij}$ | Feedback to the $k$-th block created by $U_i$ on $Chain_j$ |

After a round of new block creation, a lot of useful data are generated, including the local and global trust values of the miners who created the block, the accuracy of block creation, and so on. These data can be collected and backed up to the side chain through cross-chain interaction technology.

The trusted miners are in charge of the side chain, because they are trustworthy no matter which chain they are on. And the data can be stored in the blocks of the side chain. For example, if a miner proposes a new block, the miner's ID and chain's ID should be recorded in the block header, and the miner's trust value will be stored in block body. The history trust data of the miner can be found quickly when it is needed urgently. Generally, there are some shortcomings to storing according to the upper limit of block capacity, including slow block generation rate and low queried efficiency. So, we consider storing data according to miners. Although this will make the blockchain longer, the storage capacity of each block is small and usually less than 1 MB[37], which will not increase the overload on the sever that deploys the side chain. Moreover, the longer the blockchain is, the more difficult it is to be tampered with. Blocks can be created faster and the queried efficiency is also higher. Whenever a new block is created, the trust value of each miner will be updated accordingly, which will be transported to the side chain by cross-chain protocol and packaged into a block. We can quickly obtain the latest trust value of the miner just by searching the last block of the side chain.

This side chain can be regarded as the sharing link of historical experiences generated by users' activities on each blockchain. When it is necessary to judge whether $U_i$ on $Chain_j$ is an intensive DMB attacker, DBP can access the $U_i$'s historical experiences from the side chain through cross-chain interaction as a predictive support.

In addition, after the newly created block is broadcast to all the users on $Chain_j$, the feedback on the authenticity of the block shall be provided and uploaded by these users to the side chain. $F^k_{ij}$ is the feedback from other users to the $k$-th block created by $U_i$ on $Chain_j$, which can be used to adjust the prediction results of DBP scheme and improve the prediction

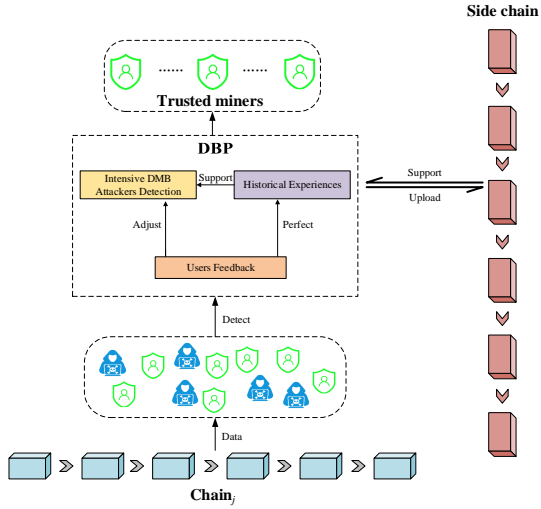

Fig. 4. Architectural overview of DBP.

accuracy.

DBP only needs a binary prediction result. Considering that the prediction result of SVM [38] only depends on the number of supporting vectors and has nothing to do with the correlation between data features, it can avoid the disaster of dimension. Moreover, SVM is highly accurate and fast in dichotomy prediction, so it can be well applied to DBP.

In the DBP scheme, SVM takes the historical experiences ($\psi^T$) as the feature space and learns in the feature space to find an optimal hyperplane by analyzing the characteristics of data. Its ultimate goal is to maximize the interval. This process can be described as:

$$\begin{cases} \psi^T x + \gamma = 0, \\ \psi^T = (lt_{ij}, gt_i, t_i, f_i, L_j, N_j, F^k_{ij}), \end{cases} \quad (9)$$

where $\gamma$ is the decision constants of SVM.

According to the user data characteristics, SVM can map each user to the multidimensional space in the form of data points, which contain two types of users separated from $\Phi_1$: trusted users and intensive DMB attackers.

The first thing is to label them respectively for SVM model training. The user type is represented by $p_i$, which can be represented as:

$$p_{ij} = \begin{cases} +1, & U_i \text{ is an intensive DMB attacker}, \\ -1, & U_i \text{ is a trusted user}. \end{cases} \quad (10)$$

Thus, the two sides of the hyperplane can be divided into two different types of users, which can be represented as.

$$\begin{cases} \psi^T x + \gamma \geq +1, & p_{ij} = +1, \\ \psi^T x + \gamma \leq -1, & p_{ij} = -1. \end{cases} \quad (11)$$

In order to find an optimal hyperplane to separate these two types of users as far as possible, SVM needs to calculate the distance ($d$) between the data points and the hyperplane. In a multidimensional space, $d$ can be calculated by as:

$$\begin{cases} d = \dfrac{|\psi^T x + \gamma|}{\|\psi\|}, \\ \|\psi\| = \sqrt{lt_{ij}^2 + gt_i^2 + t_i^2 + f_i^2 + L_j^2 + N_j^2 + (F^k_{ij})^2}. \end{cases} \quad (12)$$

By finding the data points closest to the hyperplane and taking them as the boundary of training results, SVM can separate different data points. At this point, $d$ is at the minimum value ($d_{min}$). The data point closest to the hyperplane is the support vector of SVM. The plane where the support vector is located becomes the classification decision surface. $d_{min1}$ represents the minimum distance between trusted users and the hyperplane. $d_{min2}$ is the minimum distance between intensive DMB attackers and the hyperplane. Therefore, the gap between the two types of support vector ($\Delta d$) is equal to $d_{min1}+d_{min2}$, which can be further calculated as:

$$\Delta d = \frac{|\psi^T x + \gamma + 1 - (\psi^T x + \gamma - 1)|}{\|\psi\|} = \frac{2}{\|\psi\|}. \quad (13)$$

When $\Delta d$ is at the maximum value, the accuracy of training results will be the highest. Specially, $\Delta d$ will get the maximum value ($\Delta d_{max}$) under the following constraints:

$$\begin{cases} \Delta d_{\max} = \min_{\psi, \gamma} \dfrac{1}{2} \|\psi\|^2, \\ s.t. \ p_{ij}(\psi^T x + \gamma) \geq 1, \end{cases} \quad (14)$$

where $p_{ij}(\psi^T x + \gamma) \geq 1$ is derived from equation (11).

By introducing Lagrange multiplier method, the result of equation (14) can be obtained by SVM. That is, the objective function can be expressed as:

$$L(\psi, \gamma, \alpha) = \frac{1}{2} \|\psi\|^2 + \sum_{r=1}^{s} \mu_r \left(1 - p_{ij}(\psi^T x_r + \gamma)\right), \quad (15)$$

where $\mu$ is the Lagrangian multiplier and $s$ is the number of data features. In the DBP scheme, $s=7$.

Since Lagrangian functions have the property of duality, it is usually possible to solve its duality problem to get the final result.

Let $\dfrac{\partial L}{\partial \psi} = 0, \dfrac{\partial L}{\partial \gamma} = 0$, we can obtain:

$$\psi = \sum_{r=1}^{s} \mu_r p_{ij}^r x_r,$$

$$\sum_{r=1}^{s} \mu_r p_{ij}^r = 0. \quad (16)$$

Substituting equation (16) into equation (15), we further obtain:

$$\begin{aligned} L(\psi, \gamma, \alpha) &= \frac{1}{2} \|\psi\|_2^2 + \sum_{r=1}^{s} \mu_r \left[1 - p_{ij}^k(\psi^T x_r + \gamma)\right] \\ &= \frac{1}{2} \psi^T \psi - \sum_{r=1}^{s} \mu_r p_{ij}^r \psi^T x_r - \sum_{r=1}^{s} \mu_r p_{ij}^r \gamma + \sum_{r=1}^{s} \mu_r \\ &= -\frac{1}{2} \left(\sum_{r=1}^{s} \mu_r p_{ij}^r x_r\right)^T \left(\sum_{r=1}^{s} \mu_r p_{ij}^r x_r\right) - \gamma \sum_{r=1}^{s} \mu_r p_{ij}^r + \sum_{r=1}^{s} \mu_r \quad (17) \\ &= \frac{1}{2} \sum_{r=1}^{s} \sum_{q=1}^{s} \mu_r \alpha_q p_{ij}^r p_{ij}^q (x_r x_q) \\ &\quad - \sum_{r=1}^{s} \mu_r p_{ij}^r \left(\left(\sum_{h=1}^{s} \mu_t p_{ij}^q x_q\right) \cdot x_r + \gamma\right) + \sum_{r=1}^{s} \mu_r \\ &= -\frac{1}{2} \sum_{r=1}^{s} \sum_{q=1}^{s} \mu_r \mu_q p_{ij}^r p_{ij}^q (x_r \cdot x_q) + \sum_{r=1}^{s} \mu_r, \end{aligned}$$

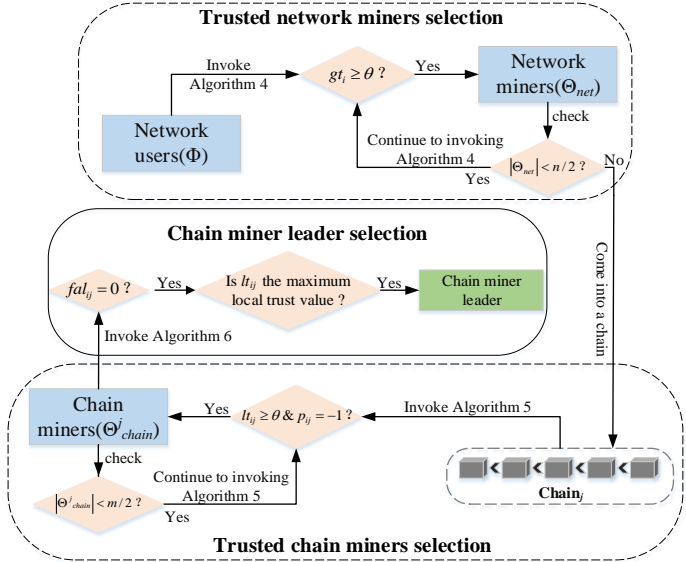

Fig. 5. The relationship of trusted miners selection

where $r$ and $q$ are the data features quantities of trusted users and intensive DMB attackers respectively.

With equation (17), we can get the Lagrangian multiplier ($\mu$). Substituting $\mu$ into equation (16), $\psi^T$ and $\gamma$ can be expressed as:

$$\psi^T = \sum_{r=1}^{s} \mu_r p_{ij}^r x_r,$$
$$\gamma = p_{ij}^q - \sum_{r=1}^{s} \mu_r p_{ij}^r (x_r \cdot x_q), \quad (18)$$

where $\psi^T$ and $\gamma$ can be used to determine the optimal hyperplane selection.

So far, the prediction model can be built by combining equation (18) into equation (15).

$$f(x) = \frac{1}{2}\|\psi\|^2 + \sum_{i=1}^{n}\left(1 - p_{ij}\left(\sum_{r=1}^{s} \mu_r p_{ij}^r x_r x_i + p_{ij}^q - \sum_{r=1}^{s} \mu_r p_{ij}^r (x_r \cdot x_q)\right)\right). \quad (19)$$

With this model, the predicted value $p_{ij}$ belonging to a user can be obtained by inputting the user data characteristics.

Let $\Phi_5$ denotes the set of intensive DMB attackers on $\text{Chain}_j$, Algorithm 3 can be performed to detect intensive DMB attackers.

### D. Trust Miners Selection in Multi-chains

Current mainstream consensus protocols generally require more than 51 percent of users to perform miner duties. At the beginning of the network, the first miners could be randomly generated. Once there is a history of behaviors, honest users should be selected as trusted miners. The relationship of trusted miners selection is shown in Fig. 5.

Algorithm 4 can be triggered to select trusted network miners ($\Theta_{net}$) from the entire network of users ($\Phi$).

Trusted network miners selection should be a dynamic updating process. Once $gt_i < \theta$, $U_i$ will be deleted from $\Theta_{net}$. When $|\Theta_{net}| < n/2$ due to the removal of unqualified miners, Algorithm 4 must be triggered again.

In our consensus scheme, only the chain miners ($\Theta_{chain}$) selected from $\Theta_{net}$ have the authority to create blocks. Take

---

**Algorithm 3** Our proposed Intensive DMB attackers detection

**Input:** $\Phi_1$, $lt_{ij}$, $gt_i$, $t_i$, $f_i$, $L_j$, $N_j$, $F^k_{ij}$, $p_{ij}$,

**Output:** $\Phi_5$

1: Acquire users' historical experiences from the side chain
2: Extract $\psi^T$, $\gamma$ from the historical experiences
3: Construct $f(x) = \frac{1}{2}\|\psi\|^2 + \sum_{r=1}^{s}\left(1 - p_{ij}(\psi^T x_r + \gamma)\right)$
4: Export the prediction model
5: **for each** $U_i \in \Phi_1$ **do**
6:     Enter ($lt_{ij}$, $gt_i$, $t_i$, $f_i$, $L_j$, $N_j$, $F^k_{ij}$) into the prediction model
7:     Get $p_{ij}$
8:     **if** $p_{ij} = +1$ **then**
9:         $\Phi_5 = \{U_i\} \cup \Phi_5$
10:    **end if**
11: **end for**

---

**Algorithm 4** Trusted network miners selection

**Input:** $\Phi$, $\Xi$

**Output:** $\Theta_{net}$

1: **for each** $U_i \in \Phi$ **do**
2:     **if** $gt_i \geq \theta$ **then**
3:         $\Theta_{net} = \{U_i\} \cup \Theta_{net}$
4:     **end if**
5: **end for**
6: **repeat**
7:     **if** $|\Theta_{net}| < n/2$ **then**
8:         Randomly select a user $U_p$ from $\Phi$
9:         $\Theta_{net} = \{U_p\} \cup \Theta_{net}$
10:    **end if**
11: **until** $|\Theta_{net}| > n/2$

---

$\text{Chain}_j$ as an example again, $\Lambda_j$ denotes the set of active users who often participate in activities along $\text{Chain}_j$. Since the number of active users is dynamic, the maximum number of active users over a period of time can be used to measure the required number of chain miners. That is, $|\Theta^j_{chain}| > m/2$.

In a round of block creation, Algorithm 5 can be executed to selected trusted chain miners ($\Theta^j_{chain}$) on $\text{Chain}_j$.

After a block is created, the chain miner leader shall be elected periodically to confirm the block validation and accept it. Algorithm 6 can be carried out to elect and update the chain miner leader ($L_j$) on $\text{Chain}_j$.

When $fal_{ij} \geq 1$ or the life time of $L_j$ is expired, Algorithm 6 must be triggered again to update the chain miner leader.

**Algorithm 5** Trusted chain miners selection on Chain$_j$

**Input:** $\Lambda_j$, $\Theta_{net}$, $LT_{n\times h}$
**Output:** $\Theta^j_{chain}$

1: **for** each $U_i \in \Lambda_j$ **do**
2:     **if** $U_i \in \Theta_{net}$ **then**
3:         Perform Algorithm 3
4:         **if** $p_{ij}== -1$ && $lt_{ij} \geq \theta$ **then**
5:             $\Theta^j_{chain} = \{U_i\} \cup \Theta^j_{chain}$
6:         **end if**
7:     **end if**
8: **end for**
9: **repeat**
10:     **if** $|\Theta^j_{chain}| < m/2$ **then**
11:         Randomly select a user $U_q$ from $\Lambda_j$
12:         $\Theta^j_{chain} = \{U_q\} \cup \Theta^j_{chain}$
13:     **end if**
14: **until** $|\Theta^j_{chain}| > m/2$

---

**Algorithm 6** Chain miner leader election

**Input:** $\Theta^j_{chain}$
**Output:** $L_j$

1: Initialize $\Theta_L = \varnothing$
2: **for** each $U_i \in \Theta^j_{chain}$ **do**
3:     **if** $fal_{ij}==0$ **then**
4:         $\Theta_L = \{U_i\} \cup \Theta_L$
5:     **end if**
6: **end for**
7: **for** each $U_i \in \Theta_L$ **do**
8:     **if** $lt_{ij}$ is the maximum local trust value from $\Theta_L$ **then**
9:         $L_j = U_i$
10:     **end if**
11: **end for**

**Complexity analysis:** The overall time complexity for algorithms is computed as follows. The algorithms are executed for n network miners, so the algorithm that acquire trust value of users takes $O(n)$. The creation of a new block by chain miners takes $O(m/2)$, and the remaining computations can be carried out in $O(1)$ time. Therefore, the overall complexity is $O(n)+O(m/2)+O(1)=O(n)+O(m/2)$.

## V. PERFORMANCE ANALYSIS

### A. Experimental Setup

Computer experiments are performed to analyze the performance of our proposed PoDT consensus scheme in Python 3.7. The experiment elements are shown in Table II.

We set up a blockchain network with ten chains, in which four chains will be randomly selected as kill-chains and the others are mask-chains. When network users are selected as miners to create blocks, they are divided into four types.

TABLE II
DESCRIPTION OF EXPERIMENT ELEMENTS

| Parameters | Description | Default |
|---|---|---|
| $N_n$ | Number of users | 1000 |
| $N_c$ | Number of chains | 10 |
| $\theta$ | Threshold of trust value | 0.5 |
| $\xi_1$ | Trust warning line | 0.1 |
| $\xi_2$ | High trust line | 0.4 |
| cycle | Number of cycles | 200 |

Trusted miners always create true blocks on all chains, while ordinary attackers always create false blocks on all chains. Normal DMB attackers only create false blocks on kill-chains, but behave well on mask-chains. Intensive DMB attackers alternately creating true or false blocks on kill-chains.

### B. Experimental results

We set up the peer-to-peer consensus processes of the blockchain network and perform five experiments to validate the effectiveness of PoDT.

The first three experiments are performed by cycle-based fashion to further analyze DMB attacks and show the performance of its two-level defense scheme including DiscTrust and DBP.

In the first experiment, we compare DMB attacks with ordinary attacks in terms of global trust value. We chose an attacker from each of the three types of malicious users. As shown in Fig. 6, the global trust value of an ordinary attacker is far below the threshold $\theta$. Thus, ordinary attackers can be easily detected by trust management. Along with 200 cycles, both the normal and intensive DMB attackers have the global trust value far above the threshold $\theta$. Therefore, common trust management systems will be harder to detect them.

To differentiate the normal and intensive DMB attackers, we randomly select two kill-chains to further observe their local trust value. As shown in Fig. 7, the local trust value of a normal DMB attacker is far below the threshold $\theta$ on kill-chains, while an intensive DMB attacker is far above the threshold $\theta$.

As we know, malicious users launch DMB attacks to increase their trust value, which may cause a lot of malicious responses at each cycle. The DMB malicious responses may result in the unnecessary waste of network resources. So, reducing these malicious responses is the best measure to suppress DMB attacks.

In the second experiment, we validate the performance of the DiscTrust scheme in reducing normal DMB malicious responses and the DBP scheme in reducing intensive DMB malicious responses, respectively.

As shown in Fig. 8, the normal DMB malicious responses of Baseline are far more than DiscTrust. Without any defense measures in the Baseline scheme, normal DMB attackers' local trust value decreases slowly, they can get more chance to launch DMB attacks, resulting in the most malicious responses. With the DiscTrust scheme, normal DMB attackers will be forbade to join the chain miners because of their lower trust value, and thereby suppressing malicious responses.

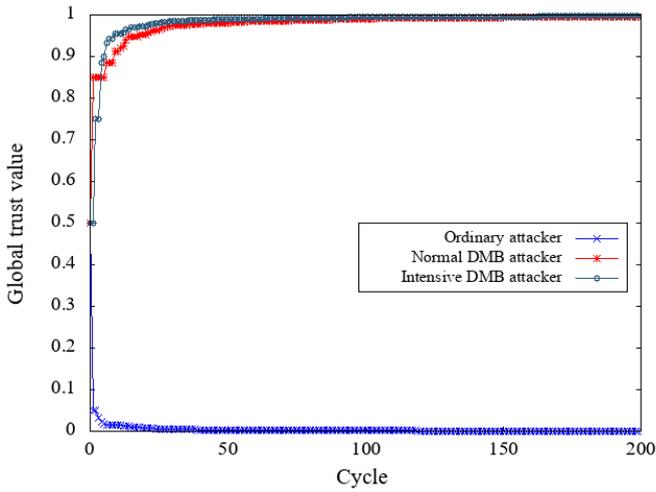

Fig. 6. Global trust value of different attackers.

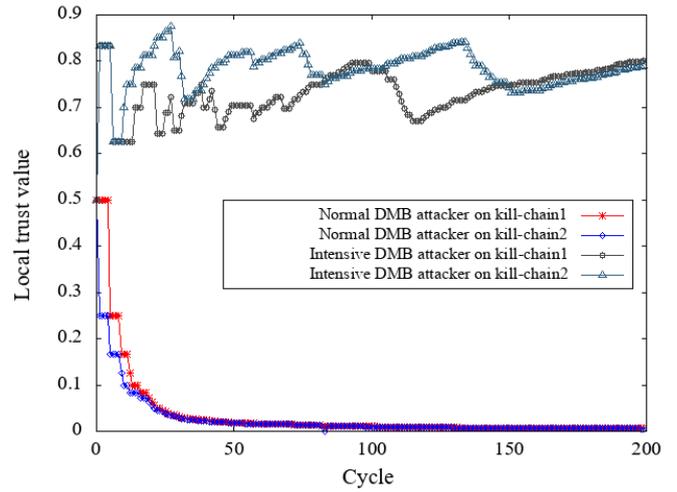

Fig. 7. Local trust value of different DMB attackers on kill-chains.

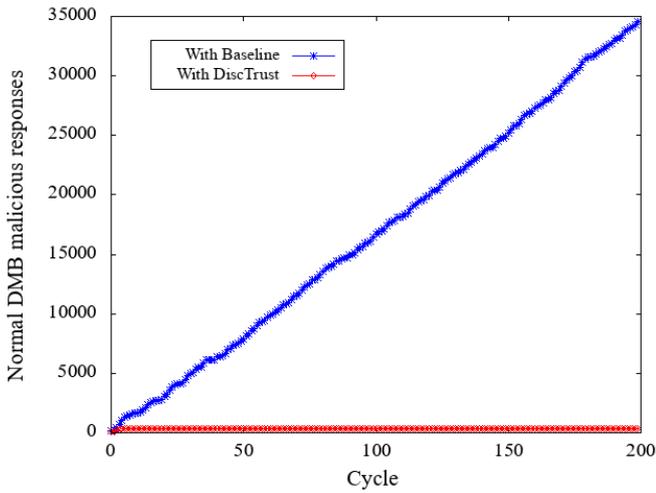

Fig. 8. Suppressing normal DMB malicious responses.

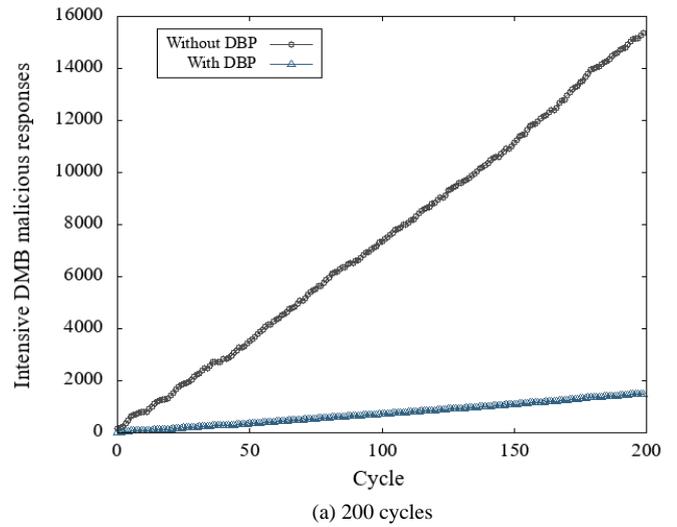

(a) 200 cycles

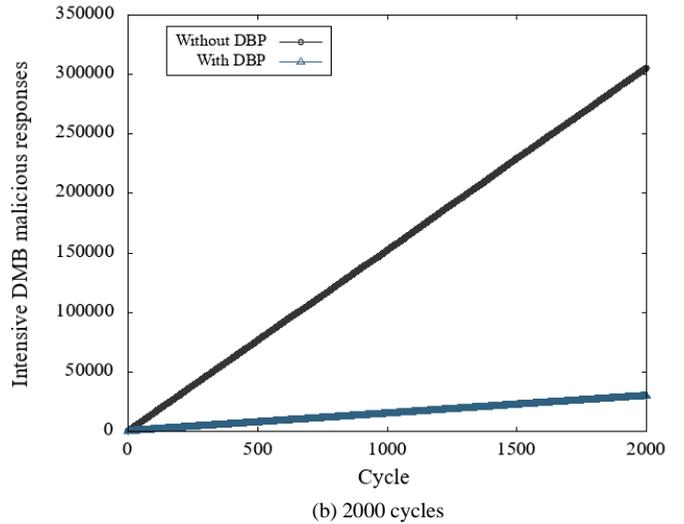

(b) 2000 cycles

Fig. 9. Suppressing intensive DMB malicious responses.

We also compare the DiscTrust scheme reinforced by DBP and the DiscTrust scheme without DBP. As shown in Fig. 9, DBP can suppress intensive DMB malicious responses effectively, since DBP can predict the attacker in advance. In order to cover sufficient design space, we have increased the number of chains to 100, and the cycles is also increased to 2000, the results show that the DBP scheme can still suppress intensive DMB attacks well.

In the third experiment, we further evaluate the performance of the two-level defense scheme in suppressing the attack success ratio. Without loss of generality, we define the ratio of false blocks amount successfully created by DMB attackers to newly created blocks amount at each cycle as the attack success ratio.

As shown in Fig. 10, the normal DMB attack success rate in DiscTrust is also lower than Baseline. Under the protection of the distinctive trust evaluation, normal DMB attackers can be effectively detected by analyzing their local trust value on kill-chains, so their attacks cannot succeed. As shown in Fig. 11, DBP can also reduce intensive DMB attack success rate effectively.

Specifically, how is the detection performance of DBP for intensive DMB attackers? With the increase of users from 1000 to 10000, the detection rate of DBP for intensive DMB attackers can generally reach more than 90%, as shown in Fig. 12. Even if the percentage of intensive DMB attackers is 50%, it can also achieve a better detection rate. Even due to the increasing of number of nodes, the size of the data set is larger. DBP scheme can analyze more data and make our detection rate higher.

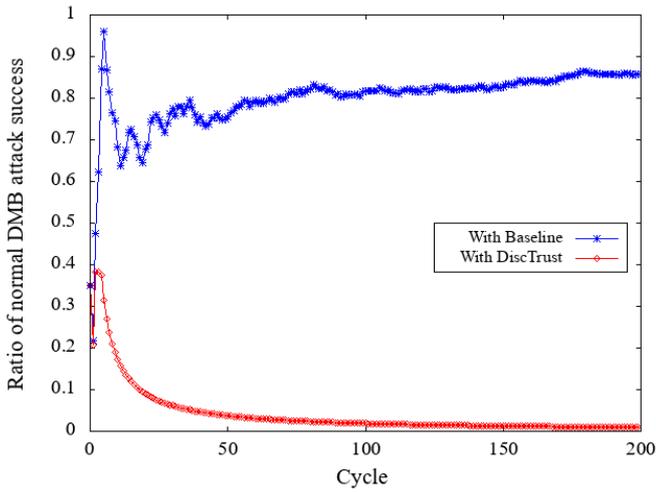

Fig. 10. Suppressing normal DMB attack success ratio.

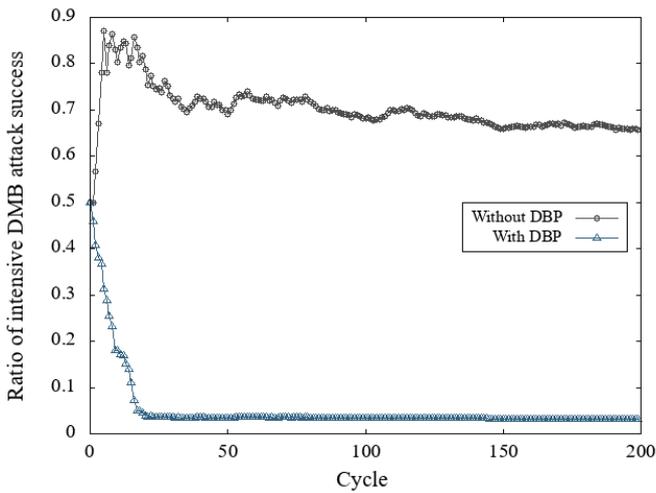

Fig. 11. Suppressing intensive DMB attack success ratio.

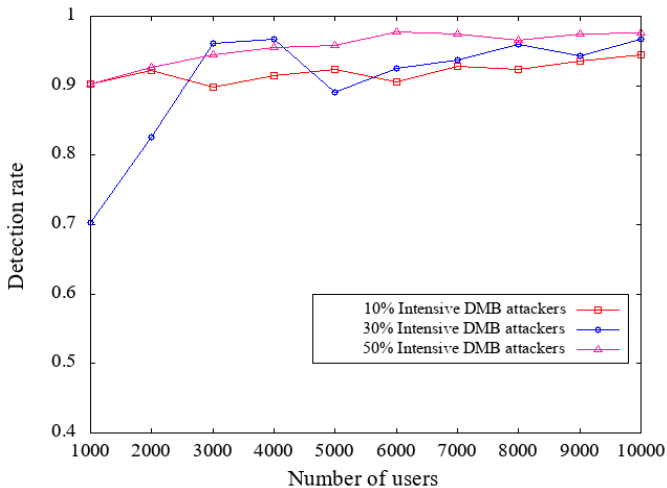

Fig. 12. The detection rate of DBP.

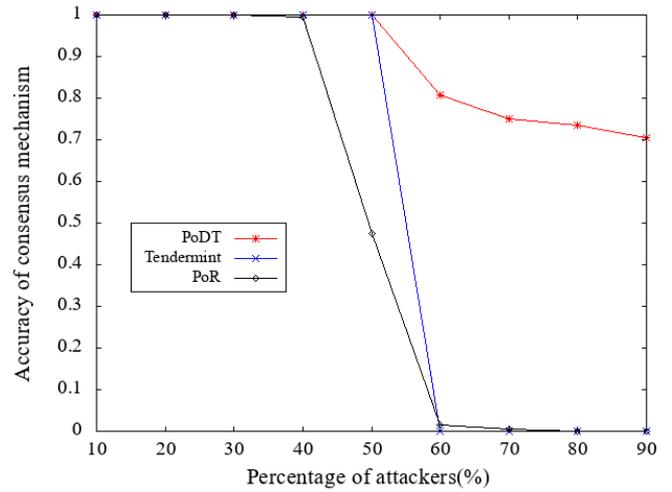

(a) Kill-chains

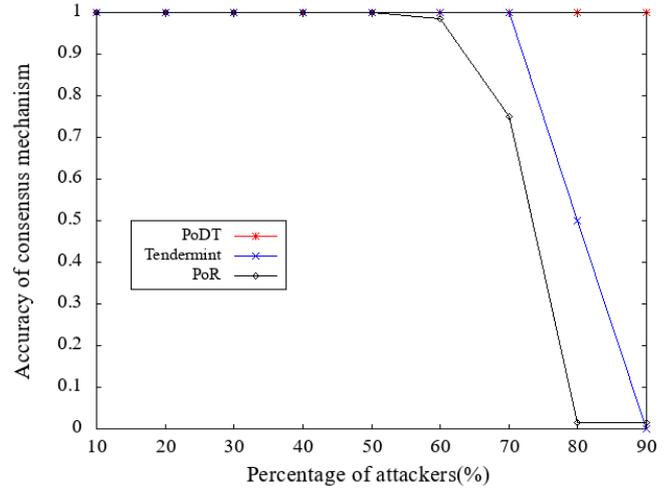

(b) Mask-chains

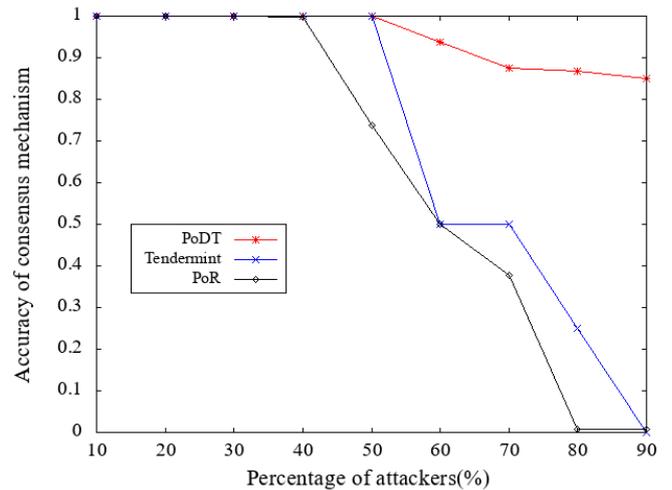

(c) Whole blockchain network

Fig. 13. Accuracy comparison of PoDT, Tendermint and PoR.

In the last three experiments, we compare PoDT with Tendermint [18] and PoR [19]. The three typical types of consensus schemes are based on the fair multi-miners participation, in which each user can fairly compete to become a miner in a blockchain network.

In the fourth experiment, we compared the accuracy of PoDT with Tendermint and PoR. The average accuracy of 200 rounds of block creation was employed as the experimental result. The attackers fall into three categories, including ordinary attackers, normal DMB attackers and intensive DMB attackers. Under a certain percentage, the number of these three kinds of attackers is randomly assigned. We vary the percentage of attackers to observe the accuracy comparison.

As shown in Fig. 13, PoDT can achieve nice accuracy. On mask-chains, the accuracy of PoDT is the best. Even though PoDT drops a little on kill-chains, it is still better than

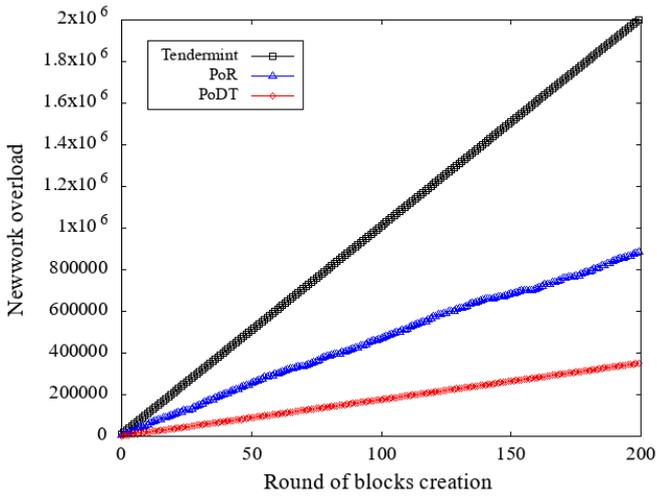

(a) 200 round of blocks creation

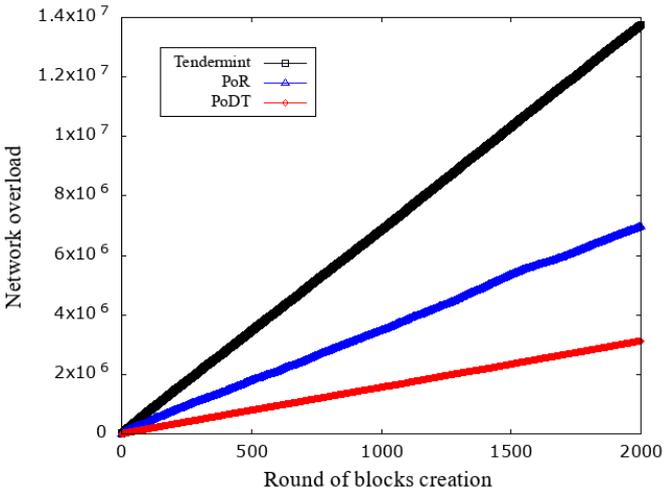

(b) 2000 round of blocks creation

Fig. 14. Network overload comparison of PoDT, Tendermint and PoR.

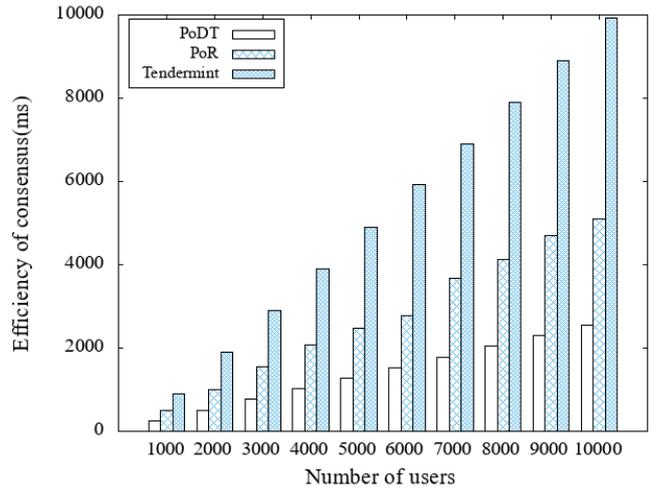

Fig. 15. Efficiency comparison of PoDT, Tendermint and PoR

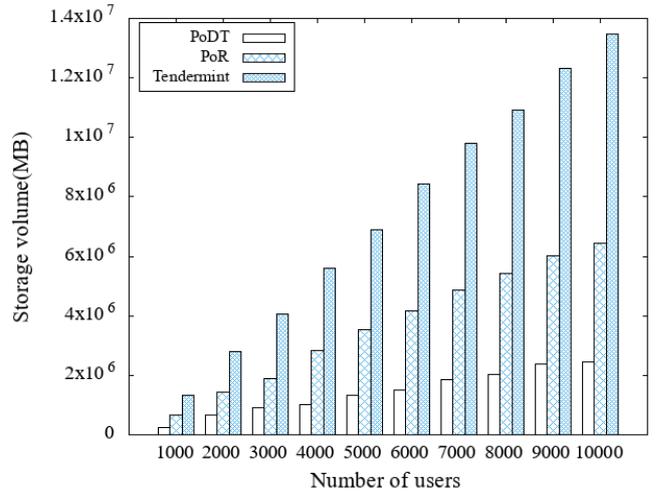

Fig. 16. Storage volume comparison of PoDT, Tendermint and PoR

Tendermint and PoR. From the perspective of the whole blockchain network, PoDT is slightly better than the others since trust management is considered to select honest miners. Although the accuracy of PoDT will drop sharply when the percentage of attackers is more than 50%, such an extreme situation would not appear in real blockchain networks.

In the fifth experiment, we vary the round of blocks to analyze the network overload of PoDT, Tendermint and PoR. As shown in Fig. 14, the network overload of Tendermint is the biggest. This is because all users of the blockchain network are selected as miners in Tendermint. In addition, we found that the random selection of chain miners from the network miners can generate lower network overload than the random selection of miners from the whole blockchain network. Similarly, when there are 100 chains and the cycle is 2000, the performance in network overload of PoDT is always better than Tendermint and PoR.

In the final experiment, in order to more intuitively evaluate the performance of PoDT, we verified the efficiency and the storage volume consumption of PoDT without loss of generality, and the size of a block is usually set to 1 MB [37]. As shown in Fig. 15 and Fig. 16, it is very easy to find that the efficiency of PoDT is better than others with the increase of users from 1000 to 10000. Because there are only a few trusted miners can participate in the consensus in PoDT, the consumption of time can be greatly reduced. Moreover, the blocks are only backed up in theses trusted nodes instead of all nodes, so PoDT has the lowest consumption in storage volume.

As a consequence, PoDT can achieve higher accuracy at a better performance.

## VI. INDUSTRIAL APPLICATIONS DISCUSSION

A blockchain network with multiple chains running together can be applied in the data sharing of some industries, such as medical, finance, electric power, telecommunications, etc. For instance, it can help build a medical data sharing alliance, so the medical "data island" can be overcome, as shown in Fig. 17. Data interconnection and sharing among medical institutions, social insurance institutions, research institutions and regulators can effectively promote the progress of medical research, medical finance, hospital management, clinical medicine, big data medicine, etc.

In this case, each chain can use the same consensus mechanism to promote the efficiency. However, this still have security risks. For example, once some nodes are compromised, it is possible that they will launch DMB attack, then we may lose control of the blockchain network. Taking into account the requirements of defending against DMB attacks, trust mechanism can be introduced to our PoDT scheme. Implementing a universal reputation score can help us quickly and effectively detect malicious miners whose trust value is below the threshold and prevent them from destroying the consensus mechanism. This is also a function that single- chain consensus cannot achieve. Considering the trust values from the network level and quantifying the behavior of miners in consensus, the trust value of miners can be divided into global trust value and local trust value. Only when both of them are above the threshold can the miner be trusted, thus avoiding a widespread damage to the blockchain in advance. What's more, we can regain the control of blockchain networks after it is lost.

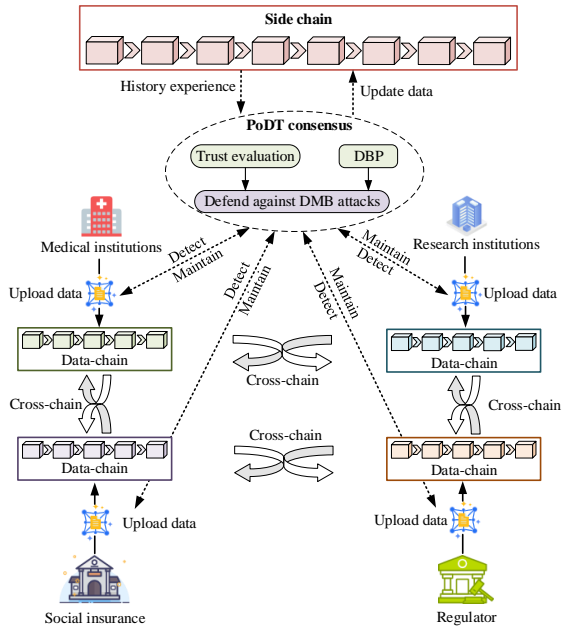

Fig. 17. Industrial application case of PoDT

In our PoDT scheme, a side chain is designed to store their trust value, which can make it as tamper-resistant. According to the trust value of miners, some trusted miners are randomly selected to participate in the consensus, which can promote the efficiency of consensus and reduce the consumption of storage volume. Moreover, our PoDT can achieve the security guarantee of blockchain network with low complexity of algorithms.

## VII. CONCLUSION AND FUTURE WORK

In this paper, we propose an advanced consensus scheme named PoDT for multi-chains to defend against DMB attacks in blockchain networks. DMB attacks can be launched from two aspects of threats: normal and intensive. With the help of local trust value, DiscTrust is introduced to detect normal DMB attackers since they always behave vicious on kill-chains. Even if intensive DMB attackers have the ability to maintain high trust on kill-chains, the DBP scheme can strengthen DiscTrust to detect them. On this basis, three algorithms can be designed to select trusted miners for multi-chains. Experimental results show that our consensus scheme is secure against DMB attacks in multi-chains environments. More importantly, experimental results also show that PoDT is more effective than Tendermint and PoR in terms of block creation.

Cross-chain technologies can connect multi-chains to form the internet of chains, which extensively expands the application of the blockchain. For future works, we will investigate the cloud-assisted internet of chains for threat intelligence sharing, where our PoDT scheme can ensure the security of multi-chains consensus.